\documentclass[prl,superscriptaddress,twocolumn,showpacs,amsmath,amssymb]{revtex4}

\usepackage{graphicx}
\usepackage{dcolumn}
\usepackage{bm}
\usepackage{color}
\usepackage{mathrsfs}

\usepackage{xcolor}
\newcommand{\red}[1]{\textcolor{red}{\bf #1}}

\begin{document}

\preprint{APS/123-QED}

\title{Evolution of density of states and spin-resolved ``checkerboard'' pattern associated with Majorana bound state}

\author{Takuto Kawakami}
\email{KAWAKAMI.Takuto@nims.go.jp}
\affiliation{International Center for Materials Nanoarchitectonics (WPI-MANA),
National Institute for Materials Science, Tsukuba 305-0044, Japan}

\author{Xiao Hu}
\email[Correspondence to: ]{HU.Xiao@nims.go.jp}
\affiliation{International Center for Materials Nanoarchitectonics (WPI-MANA),
National Institute for Materials Science, Tsukuba 305-0044, Japan}

\date{\today}

\begin{abstract} {
In terms of Bogoliubov-de Gennes approach, we investigate Majorana bound state (MBS) in vortex of proximity-induced superconductivity on the surface of topological insulator (TI).
Mapping out the local density of states (LDOS) of quasiparticle excitations as a function of energy and distance from vortex center, it is found that the spectral
distribution evolves from ``V''-shape to ``Y''-shape with emergence of MBS upon variation of chemical potential, consistent with the STM/STS measurement in a very recent experiment [Xu {\it et al.}, Phys. Rev. Lett. {\bf 114}, 017001 (2015)] on Bi$_2$Te$_3$ thin layer on the top of NbSe$_2$.
Moreover, we demonstrate that there is a ``checkerboard'' pattern in the relative LDOS between spin up and down channels, which maps out directly the quantum mechanical
wave function of MBS. Therefore, spin-resolved STM/STS technique is expected to be able to provide phase sensitive evidence for MBS in vortex core of topological superconductor.}
\end{abstract}


\maketitle

{\it Introduction.---}
Majorana bound states (MBSs) are under intensive search in topolgoical superconductors
(SCs)~\cite{hasan2010, qi2011}.
Because of the peculiar property that a particle is equivalent to its antiparticle~\cite{majorana},
non-Abelian quantum statistics can be generated which is believed to be potentially important for achieving decoherence-free topological quantum computation~\cite{ivanov2001,nayak2008,alicea2011, liang2012,wu2014,alicearpp,beenakkerarcp}.
MBSs were first predicted in the spinless pairing states of fermions including the Pfaffian state of quantum Hall system with $5/2$ filling~\cite{read2000} and ultracold atomic gases with $p$-wave Feshbach resonance~\cite{gurarie2005}.
Recently, heterostructures made of $s$-wave SC and topological insulator (TI)~\cite{fu2008},
or the combination of semiconductor with Rashba spin-orbit coupling (SOC) and
ferromagnetic insulator~\cite{sato2009,lutchyn2010,sau2010}, are proposed for realizing topological SC with the spin degree of freedom suppressed by SOC. Sr$_2$RuO$_4$~\cite{maeno2012}, UPt$_3$~\cite{tsutsumi2013} and doped TI Cu$_x$Bi$_2$Se$_3$~\cite{sasaki2012} are also discussed as possible candidates.

MBS is a unique quasiparticle excitation of topological SC carrying zero energy and zero
total angular momentum, which
appears at ends of one-dimensional systems or at vortex cores, where the
SC gap is closed~\cite{fu2008, sato2009, lutchyn2010, sau2010, kitaev2001, sau2010prb, chamon2010, cheng2010, hosur2011, zzli2014, chiu2011, hui2015, jli2014}.
The unique property of equivalence between particle and antiparticle, which renders MBS noble for advanced applications, makes them difficult to be identified
~\cite{fu2009,akhmerov2009,law2009,alicearpp}. The hurdle has been tackled experimentally.
Up to this moment, there are already several reports on possible MBSs in 1D nanowires~\cite{mourik2012, deng2012, finck2013, nadj-perge2014}, where quasiparticle
states are observed at ends of nanowires at zero-energy bias.

Very recently scanning tunneling microscopy and spectroscopy (STM/STS) experiments have been performed in vortex state of proximity-induced SC in Bi$_2$Te$_3$ thin layer~\cite{xu2015}. Mapping
the local density of states (LDOS) measured by differential conductance $dI/dV$ as a function of bias voltage and the distance from vortex core, an evolution from ``V''-shape for thin TI to ``Y''-shape for thick TI was observed, and it is conjectured that the ``Y''-shaped LDOS with its center at zero-energy bias is due to the appearance of MBSs.

\begin{figure}[t]
\includegraphics[width=80mm]{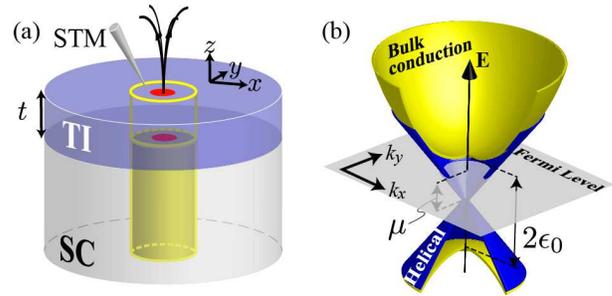}
\caption{ Schematics for the model system in the present study. (a) System geometry with a topological insulator (TI) slab on top of superconductor (SC)
carrying on one vortex, with the red dots for Majorana bound states. (b) Dispersion relation of TI with {helical} surface states
(in blue color) and the bulk valence and conduction bands (in gold color) with a gap of
$2\epsilon_0$.  The chemical potential $\mu$ is measured from the Dirac point of the topological surface states.
}
\label{fig:model}
\end{figure}

In this work, we examine the same geometry as in the experiment~\cite{xu2015} (see Fig.~\ref{fig:model})
in terms of Bogoliubov-de Gennes (BdG) analysis based on a model system including explicitly
the topological surface states of TI slab with proximity-induced SC. We clarify that, due to the relation among energy, angular momentum and spatial distribution of quasiparticle excitation, the spatial-energy distribution of LDOS evolves from ``V''-shape to ``Y''-shape corresponding to the absence and presence of MBS at vortex core when the chemical potential is reduced, which is in nice agreement with the STM/STS measurement~\cite{xu2015}. Moreover, exploring the total angular momentum of quasiparticle excitation contributed from spin and orbital angular momenta and the phase winding of vortex, we demonstrate that there should be a ``checkerboard'' pattern in spin-resolved LDOS associated with MBS. This suggests a phase sensitive way for identifying MBS as a single quasiparticle by spin-resolved STM/STS technique.

\vspace{3mm}
{\it Model Hamiltonian and BdG approach}.---
As can be read from Fig.~4 of Ref.~\cite{xu2015}, when the thickness of TI slab increases,
the Fermi level shifts from a position cutting the bulk conduction band [for one to four quintuple layers (QLs)] to a position
below the conduction band (for five and six QLs), while the size of band gap between the bulk conduction
and valance
bands remains almost unchanged. For the whole range of thickness, the SC gap can be taken as constant as a fairly good approximation (see Fig.~2 of Ref.~\cite{xu2015}). In order to
model the situation realized in the experiment, we consider a system of TI slab carrying on topological
surface states and with proximity-induced SC as shown schematically in Fig.~\ref{fig:model}(a),
where the chemical potential is varied while the thickness of TI slab and the SC gap are fixed.
The BdG Hamiltonian is given by
\begin{eqnarray}\label{eq:bdg}
\mathcal{H}_{\mathrm{BdG}} = \left(\begin{array}{cc}
\hat{\mathcal{H}}_{\mathrm{TI}}(\bm{r}) & \hat{\Delta}(\bm{r}) \\
\hat{\Delta}^{\dag}(\bm{r}) & -\hat{\mathcal{H}}_{\mathrm{TI}}^{\ast}(\bm{r})
\end{array}\right)
\end{eqnarray}
with the effective Hamiltonian of TI
\begin{eqnarray}
\hat{\mathcal{H}}_{\mathrm{TI}} = \epsilon\hat{\sigma}_z -i v_{\mathrm{F}}\left[e^{-i\phi\hat{s}_z}\left(\partial_r\hat{s}_x + \frac{1}{r}\partial_\phi\hat{s}_y\right) + \partial_z\hat{s}_z\right]\hat{\sigma}_x -\mu.\nonumber
\end{eqnarray}
and Dirac mass
\begin{equation}
\epsilon = -\epsilon_0 - \frac{1}{2m}\left(\partial_r^2 + \frac{1}{r}\partial_r+\frac{1}{r^{2}}\partial_\phi^2+ \partial_z^2\right)
\end{equation}
in the cylindric coordinates,
where $\hat{s}_i$ and  $\hat{\sigma}_i$ are Pauli matrices for spin up and down states and
orbitals of Bi and Te with even and odd parities respectively.
The bulk gap between valance and conduction bands is taken as $2\epsilon_0$ and Fermi level $\mu$ is measured from the Dirac point of the topological surface states of TI;
$\mu>\epsilon_0$ corresponds to thin TI slabs and $\mu \lesssim\epsilon_0$ to
thick TI slabs in the experiment~\cite{xu2015}.
Since the energy gap of Bi$_2$Te$_3$ is $\epsilon_0\simeq 0.1$eV and the proximity-induced
SC gap is $\Delta_0\simeq 1$ meV in the experiment~\cite{xu2015}, we take $\Delta_0/\epsilon_0=0.02$
for numerical calculation in the present work, noticing that this value influences substantially
the distribution of spectral weight. The Fermi velocity $v_{\mathrm{F}}$ of topological surface state corresponds
the slope of linear dispersion in Fig.~\ref{fig:model}(b), which yields the unit for lateral length $k_0^{-1}=v_{\mathrm{F}}/\epsilon_0$.
Taking $v_{\rm F}\simeq 0.2$ nm$\cdot$eV for Bi$_2$Te$_3$ from another experiment~\cite{chen2009}, one has $k_0^{-1}\simeq 2$nm.
The effective mass is taken as $m=2\epsilon_0/v_{\rm F}^2$ for simplicity.

\begin{figure}[t]
 \includegraphics[width=80mm]{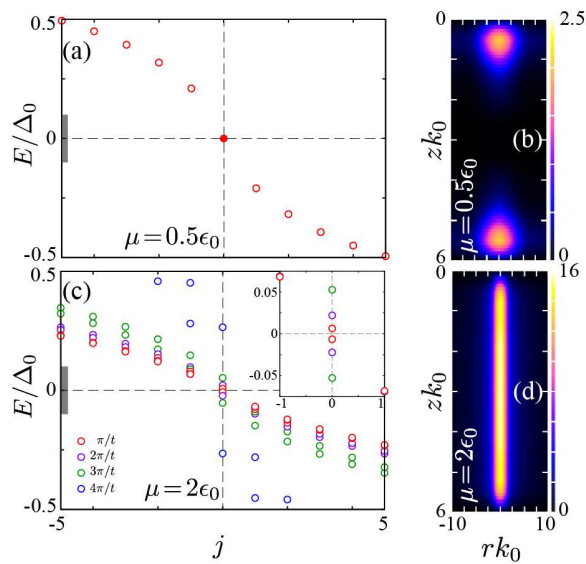}
 \caption{(a) and (c) Eigen energy of quasiparticle excitation in vortex as a function of total angular moment $j$.
 The red dot in (a) indicates two degenerate MBSs localized at the top and bottom surface of TI slab. Colors in (c) are for values of momentum along $z$ direction.
 (b) and (d) LDOS in a window close to zero energy $|E|/\epsilon_0<0.002$  as denoted by the gray bar in (a) and (c). Parameters are $\Delta_0/\epsilon_0=0.02$ and $t=6 k^{-1}_0$.
 }
\label{fig:EandZDOS}
\end{figure}

For simplicity we consider the case with only one vortex, where the
SC gap takes the form $\hat{\Delta}=e^{i\phi}\Delta_0 i\hat{s}_y$ with $\phi$ the azimuthal angle in the cylindrical coordinate.
The eigen wave functions of quasiparticle excitations
take the form
\begin{eqnarray}
\left(\begin{array}{c}
\vec{u}_{j,E}^{s}(\bm{r}) \\
\vec{v}_{j,E}^{s}(\bm{r})
\end{array}\right) =
\left(\begin{array}{c}
e^{i({j-\frac{s}{2}+\frac{1}{2}})\phi} \vec{U}_{{j},E}^{s}(r,z)\\
e^{i({j+\frac{s}{2}-\frac{1}{2}})\phi} \vec{V}_{{j},E}^{s}(r,z)
\end{array}\right),\label{eq:uv}
\end{eqnarray}
where $\vec{u}_{j,E}^{s}$, $\vec{v}_{j,E}^{s}$, $\vec{U}_{j,E}^{s}$ and $\vec{V}_{j,E}^{s}$, the electron and hole wave functions, are two-component vectors referring to the two orbitals, and $j$ and $E$ are the angular momentum and eigen energy.
For numerical diagonalization, we expand the wave functions $\vec{U}_{{j},\nu}^{s}(\bm{r})$ and $\vec{V}_{{j},\nu}^{s}(\bm{r})$ in terms of the Bessel orthonormal functions in radial
direction~\cite{gygi1991} and the Gauss-Lobatto functions in $z$ direction~\cite{mizushima2010}.
Since the coherence length $\xi=v_{\rm F}/\Delta_0=50k_0^{-1}$ is large, a radius $R=250k_0^{-1}$ is used for numerical calculations.

When the Fermi level lies in between the bulk conduction and valence bands, namely $\mu <\epsilon_0$ (thick TI slab in experiments),
SC gap is opened only in the surface Dirac dispersions of TI. As the result, two MBSs with
zero energy and zero angular momentum appear at the top and bottom surfaces of TI slab as shown in  Fig.~\ref{fig:EandZDOS}. The energy difference between
MBSs and the lowest excitations is $\delta E\simeq 0.2\Delta_0(\simeq 5\Delta^2_0/\mu)$ for $\mu/\epsilon=0.5$.
The spatial distributions of the two MBSs on the top and bottom surfaces are the same since
in the present simplified model the effect of SC substrate is taken into account by a uniform
proximity-induced SC gap.
Because the two MBSs are localized at the two surfaces of TI slab, they are well described by the 2D model~\cite{fu2008},
and the analytical solution~\cite{chamon2010, cheng2010}.

When the Fermi level falls into the bulk conduction band, namely $\mu>\epsilon_0$ (thin TI slab in experiments),
low-energy quasiparticle excitations carrying finite momenta $k_z=n\pi/t$ with $t$ the thickness of TI slab
spread along $z$ direction in vortex core as can be seen in Figs.~\ref{fig:EandZDOS}(c) and (d).
Due to these states in the normal vortex core
the two MBSs on surfaces interact with each other~\cite{hosur2011} yielding a mini gap
as can be seen in inset of Fig.~\ref{fig:EandZDOS}(c).

\begin{figure}[t]
\includegraphics[width=80mm]{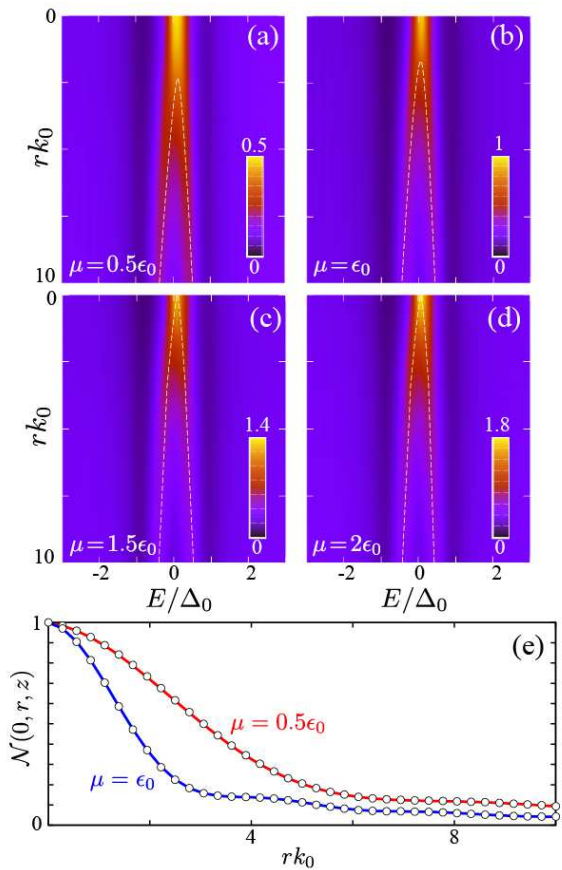}
\caption{(a)-(d) LDOS as a function of radius $r$ and energy $E$ with smearing factor $\eta/\epsilon_0=0.004$.
The spectra are taken at $z=0.24k_0^{-1}$ from the surface of the TI slab. Dotted curves are for eye guide.
(e) Spatial distribution of MBS, with open circles for numerical results and solid curves for analytical results
 $|\vec{u}_{0,0}^{\uparrow}|=e^{-r/\xi}J_0(rk_{\rm F})$ and $|\vec{u}_{0,0}^{\downarrow}|=e^{-r/\xi}J_1(rk_{\rm F})$~\cite{chamon2010, cheng2010} where $\xi=\Delta_0/v_{\rm F}=50k_0^{-1}$.
 All parameters are the same as in Fig.~\ref{fig:EandZDOS}.
}
\label{fig:VY}
\end{figure}

LDOS for quasiparticle excitations are given by the electron wave functions,
\begin{equation}\label{eqn:dos}
\mathcal{N}(E,r,z)=\mathcal{N}_{\uparrow}(E,r,z) + \mathcal{N}_{\downarrow}(E,r,z)
\end{equation}
with
\begin{equation}\label{eqn:sdos}
\mathcal{N}_{s}(E,r,z) = \sum_{E',j}\left|\vec{u}_{{j},E'}^{s}(\bm{r})\right|^2\delta(E-E'),
\end{equation}
which is measured directly by the differential conductance $dI/dV$ in STM/STS experiments.
Judging from the continuous spectra observed in Fig.~2 in Ref.~\cite{xu2015}, the energy resolution in experiments
is larger than the energy differences between quasiparticle levels. In order to simulate this situation,
we replace $\delta(x)$ with a smearing function $C(x,\eta)=x/[\pi(\eta^2+x^2)]$ with $\eta/\epsilon_0=0.004$ in Eq.~(\ref{eqn:sdos}).

The spatial distribution of a quasiparticle excitation is mainly governed by the orbital angular momentum, which appears in the wave functions approximately given by Bessel
functions. The energy dispersion with respect to the total angular momentum then gives a correlation between energy and spatial position of maximal absolute
value of quasiparticle wave function. As a general rule, quasiparticles with higher energy locate at a position farer away from the center of vortex.
As shown in Fig.~\ref{fig:VY}, for $\mu/\epsilon_0=0.5$ and 1 (thick TI slab in experiments), accompanied by MBSs with zero energy locating around the center of vortex core
[see Figs.~\ref{fig:EandZDOS}(a) and (b)], the LDOS takes a ``Y''-shape;
for $\mu/\epsilon_0=2$ (thin TI slab in experiments), the LDOS takes a ``V''-shape corresponding to the absence of spectral weight at zero energy [see Figs.~\ref{fig:EandZDOS}(c) and (d)].
This tendency can also be seen directly from the wave function of MBSs. As shown in Fig.~\ref{fig:VY}(e), the MBSs distribute over a larger area in the vortex core when the chemical potential is lowered, since their wave function is given by the Bessel function
$J_0(rk_{\rm F})$~\cite{chamon2010,cheng2010}, where $k_{\rm F}$, the intercept of Dirac dispersion with Fermi level,
decreases with decreasing $\mu$ upon lowering the Fermi level [see Fig.~\ref{fig:model}(b)].
The above evolution of spectral shape of LDOS is consistent with the tendency observed
in the STM/STS experiments where the thickness of TI slab is changed systematically~\cite{xu2015}, and supports the existence of MBSs in thick TI slabs.

LDOS presented in a previous work exhibits asymmetry in spectrum with respect to the zero energy~\cite{zzli2014}.
The apparent difference between them and the ones in Fig.~\ref{fig:VY} is due to that a large SC gap
$\Delta_0\simeq 10$ meV was taken in that work as compared with a small smearing factor.

\begin{figure}[t]
\includegraphics[width=80mm]{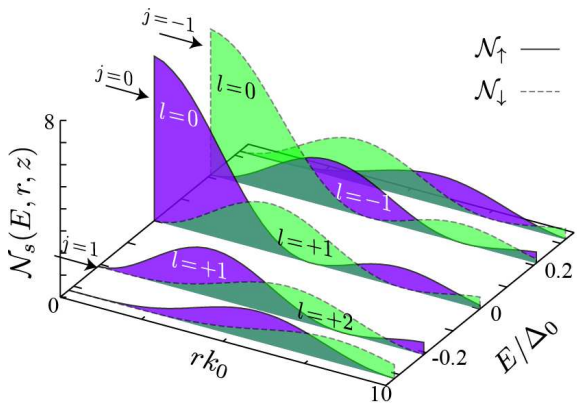}
\caption{ Wave functions of several low-energy quasiparticle excitations in vortex of topological superconductor.
Parameters are the same as Fig.~\ref{fig:EandZDOS} except for $z$=$0.24 k^{-1}_0$ and $\mu/\epsilon_0 =0.5$.
}
\label{fig:wf}
\end{figure}

\vspace{3mm}
{\it Spin-resolved LDOS.---}MBSs as quasiparticle excitation with zero energy and zero angular momentum can be better identified by spin-resolved LDOS even with the
same STM/STS resolution. To demonstrate this idea, we check the wave functions of quasiparticle excitations paying attention to spin degree of freedom.
Due to the rotational symmetry, the total angular momentum $j$
\begin{equation}
j= l +  s/2 - 1/2
\label{eqn:tam}
\end{equation}
contributed from orbital and spin angular momenta and the phase winding of SC gap
(presumed as anti-clockwise) is conserved. Therefore, it is clear that the MBS with zero total angular momentum has two components
with $l=0$ for up spin ($s=1$) and $l=1$ for down spin ($s=-1$). The detailed wave functions of MBS are displayed
in Fig.~\ref{fig:wf} with spin up and down separately, where the amplitudes oscillate with distance from vortex center in terms of Bessel functions~\cite{chamon2010, cheng2010}.
According to our numerical calculations, the excited state with energy $E=0.2\Delta_0$ and total angular momentum $j=-1$ exhibits a spatial distribution almost the same
as the MBS [see Fig.~\ref{fig:wf}].
It is intriguing to notice that, however, the oscillations in the two wave functions are out of phase when the spin state is specified. The same out-of-phase oscillations happen in
the excited state with energy $E=-0.2\Delta_0$ and total angular momentum $j=1$.

\begin{figure}[tb]
\includegraphics[width=80mm]{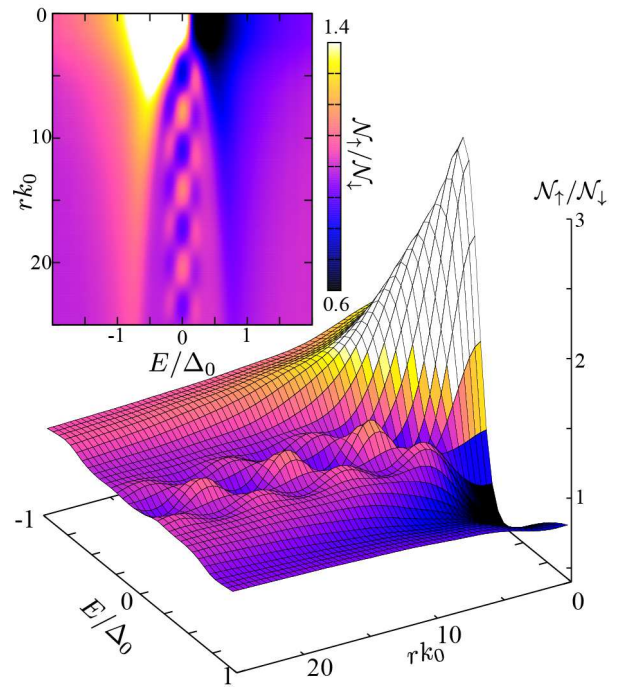}
\caption{ Relative spin-resolved LDOS $\mathcal{N}_{\uparrow}/\mathcal{N}_{\downarrow}$ as function of energy and distance from vortex center with
the same smearing parameter as in Fig.~\ref{fig:VY}. Parameters are the same as Fig.~\ref{fig:wf}.
}
\label{fig:cb}
\end{figure}

This property can be used for distinguishing MBSs from the first excited states. In order to see this explicitly, we calculate
the spin-resolved LDOSs $\mathcal{N}_{\uparrow}$ and $\mathcal{N}_{\downarrow}$ given in Eq.~(\ref{eqn:sdos}) and evaluate the relative LDOS by taking the ratio between them.
The result thus obtained for $\mu/\epsilon_0=0.5$ and $\Delta_0/\epsilon_0=0.02$
is displayed in Fig.~\ref{fig:cb}, where a checkerboard pattern is observed in the spatial-energy
mapping of LDOS with the period of 10 nm in space and 0.4 meV in energy.
We emphasize that, while the smearing factor in Figs.~\ref{fig:VY} and \ref{fig:cb} is the same, corresponding to the same STM/STS resolution, the relative spin-specific LDOS
resolves the wave function of MBS up to the quantum limit.
In a sharp contrast, the spectrum of quasiparticle excitations for $\mu/\epsilon_0=2$ is continuous, since bound states with $k_z\neq0$ spread out in $z$ direction in the vortex core,
and the relative spin-resolved LDOS $\mathcal{N}_{\uparrow}/\mathcal{N}_{\downarrow}$ loses the oscillating phase. In an anti-vortex with clockwise phase winding, the last term
in Eq.~(\ref{eqn:tam}) should change sign, and it is easy to see that the checkerboard pattern  in Fig.~\ref{fig:cb} should switch the purple-blue contrast. Therefore, the
checkerboard pattern of spin-resolved LDOS for quasiparticle excitation displayed in Fig.~\ref{fig:cb}
provides a phase sensitive evidence for MBS in vortex core of topological SC. It is worth noticing that, if the energy resolution of STM/STS measurement is high enough,
checkerboard patterns can be detected even in individual spin-resolved LDOS
(without taking the ratio between signals in two spin channels, see Appendix). 
Experimentally, spin-resolved LDOS can be obtained by using a spin-polarized STM tip~\cite{bode2003}.
It is noticed that, with very high resolution splitting in spectral weight due to spin states can be observed even in the total LDOS~\cite{hanaguri2014}.

\vspace{3mm}
{\it Conclusions.---}
We clarify that, in a vortex core of proximity-induced SC on the surface of TI, the spatial-energy mapping of local density of states of quasiparticle excitations
evolves from ``V''- to ``Y''-shape with emergence of Majorana bound state, which is in good agreement with a very
recent STM/STS experiment. Moreover, we demonstrate that the relative
density of states between spin up and down channels exhibits a checkerboard pattern reflecting the out of phase oscillations in the spin-specific wave functions of Mojarana bound
state and excited states. It provides a clue toward observing Majorana bound state in terms of phase sensitive signal by using spin-resolved STM/STS experiments.

The authors are grateful to J.-F. Jia, F.-C. Zhang, C.-H. Chen, S.-H. Pan, and T. Hanaguri for stimulating discussions.
This work was supported by the WPI Initiative on Materials Nanoarchitectonics, Ministry of Education, Culture, Sports, Science and Technology of Japan.

\appendix
\section{Appendix: LDOS in individual spin channels}

\begin{figure}[h]
\begin{center}
\includegraphics[width=80mm]{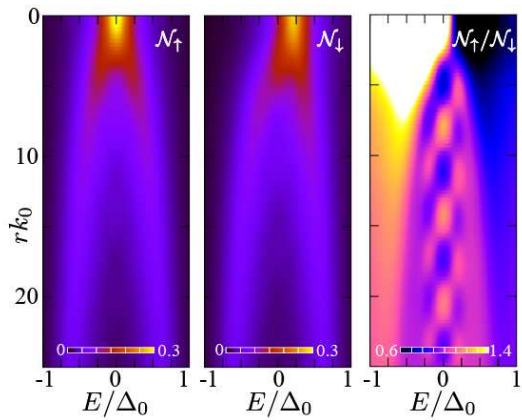}
\caption{LDOS in the two spin channels as a function of radius $r$ and energy $E$. Parameters are the
same as in Fig.~\ref{fig:cb}. For comparison, the relative spin-resolved LDOS is displayed which
is the central part of that in Fig.~\ref{fig:cb}.} \label{fig:SM1}
\end{center}
\end{figure}

Spin-resolved local densities of states (LDOS) shown in Fig.~\ref{fig:cb} are re-plotted separately
in two spin channels in Fig.~\ref{fig:SM1}. Checkerboard observed in the relative spin-resolved
LDOS cannot be seen clearly in the two individual spin-resolved LDOS, which demonstrates the advantage 
of taking their ratio in highlighting the oscillating behavior in LDOS associated with the Majorana
bound state. The LDOS is symmetric (asymmetric) in the spin-up (-down) channel corresponding to the
anti-clockwise phase winding of vortex, as can be understood from the wave functions exhibited in
Fig.~\ref{fig:wf}.

\begin{figure}[h]
\begin{center}
\includegraphics[width=80mm]{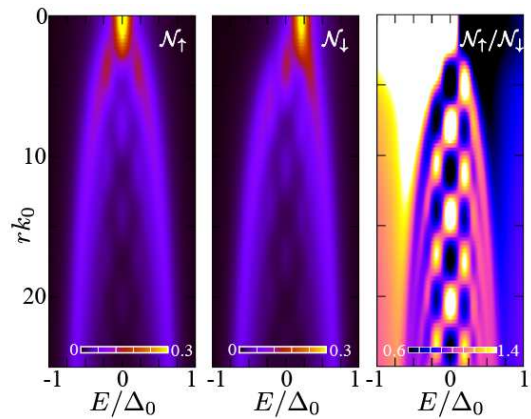}
\caption{Same as Fig.~\ref{fig:SM1} except for the smearing factor $\eta=0.002\epsilon_0$.} \label{fig:SM2}
\end{center}
\end{figure}

Meanwhile, it is worth noticing that checkerboard patterns can be identified in LDOS in individual
spin channels provided the energy resolution of STM/STS measurement is high enough. In order to demonstrate
this point, we simulate a case with smearing factor $\eta/\epsilon_0=0.002$ smaller than the one for
Fig.~\ref{fig:SM1} ($\eta/\epsilon_0=0.004$) and display the results
in Fig.~\ref{fig:SM2}. Here, once again checkerboard pattern is enhanced in relative LDOS.

\begin{thebibliography}{99}

\bibitem{hasan2010}
M. Z. Hasan and C. L. Kane, Rev. Mod. Phys. {\bf 82}, 3045 (2010).
\bibitem{qi2011} 
X.-L. Qi and S.-C. Zhang, Rev. Mod. Phys. {\bf 83}, 1057 (2011).
\bibitem{majorana} 
F. Wilczek, Nat. Phys. {\bf 5}, 614 (2009).
\bibitem{nayak2008}
C. Nayak, S. H. Simon, A.Stern, M. Freedman, and S. Das Sarma, Rev. Mod. Phys. {\bf 80}, 1083 (2008).
\bibitem{ivanov2001} 
D. A. Ivanov, Phys. Rev. Lett. {\bf 86}, 268 (2001).
\bibitem{alicea2011}
J. Alicea, Y. Oreg, G. Refael, F. von Oppen, and M. P. A. Fisher, Nature Physics {\bf 7}, 412 (2011).
\bibitem{liang2012} 
Q.-F. Liang, Z. Wang, and X. Hu, Europhys. Lett. {\bf 99}, 50004 (2012).
\bibitem{wu2014} 
L.-H. Wu, Q.-F. Liang, and X. Hu, Sci. Technol. Adv. Mater. {\bf 15} 064402 (2014).
\bibitem{alicearpp} 
J. Alicea, Rep. Prog. Phys. {\bf 75}, 076501 (2012).
\bibitem{beenakkerarcp} 
C. W. J. Beenakker, Annu. Rev. Condens. Mattter Phys. {\bf 4}, 113 (2013).
\bibitem{read2000} 
N. Read and D. Green, Phys. Rev. B {\bf 61}, 66151 (2000).
\bibitem{gurarie2005}
V. Gurarie, L. Radzihovsky, and A. V. Andreev, Phys. Rev. Lett. {\bf 94}, 230403 (2005).
\bibitem{fu2008}
L. Fu and C. L. Kane, Phys. Rev. Lett. {\bf 100}, 96407 (2008).
\bibitem{sato2009} %
M. Sato, Y. Takahashi, and S. Fujimoto, Phys. Rev. Lett. {\bf 103}, 20401 (2009).
\bibitem{lutchyn2010}
R. M. Lutchyn, J. D. Sau, and S. Das Sarma Phys. Rev. Lett. {\bf 105}, 077001 (2010).
\bibitem{sau2010}
J. D. Sau, R. M. Lutchyn, S. Tewari, and S. Das Sarma, Phys. Rev. Lett. {\bf 104}, 040502 (2010).
\bibitem{maeno2012}
Y. Maeno, S. Kittaka, T. Nomura, S. Yonezawa, K. Ishida, J. Phys. Soc. Jpn. {\bf 81}, 011009 (2012)
\bibitem{tsutsumi2013}
Y. Tsutsumi, M. Ishikawa, T. Kawakami, T. Mizushima, M. Sato, M. Ichioka, and K. Machida, J. Phys. Soc. Jpn. {\bf 82}, 113707 (2013).
\bibitem{sasaki2012}
S. Sasaki, Z. Ren, A. A. Taskin, K. Segawa, L. Fu, and Y. Ando, Phys. Rev. Lett. {\bf 109}, 217004 (2012).
\bibitem{kitaev2001}
A. Y. Kitaev, Phys. Usp. {\bf 44}, 131 (2001).
\bibitem{sau2010prb}
J. D. Sau, R. M. Lutchyn, S. Tewari, and S. Das Sarma, Phys. Rev. B 82, 094522 (2010).
\bibitem{chiu2011}
C. K. Chiu, M. J. Gilbert, and T. L. Hughes, Phys. Rev. B {\bf 84}, 144507 (2011).
\bibitem{zzli2014}
Z.-Z. Li, F.-C. Zhang, and Q.-H. Wang, Sci. Rep. {\bf 4}, 6363 (2014).
\bibitem{hosur2011}
P. Hosur, P. Ghaemi, R. S. K. Mong, and A. Vishwanath, Phys. Rev. Lett. {\bf 107}, 097001 (2011).
\bibitem{chamon2010}
C. Chamon, R. Jackiw, Y. Nishida, S.-Y. Pi, and L. Santos, Phys. Rev. B {\bf 81}, 224515 (2010).
\bibitem{cheng2010}
M. Cheng, R. M. Lutchyn, V. Galitski, and S. Das Sarma, {\bf 82}, 094504 (2010).
\bibitem{jli2014}
J. Li, H. Chen, I. K. Drozdov, A. Yazdani, B. A. Bernevig, and A. H. MacDonald
Phys. Rev. B {\bf 90}, 235433 (2014).
\bibitem{hui2015}
H.-Y Hui, P. M. R. Brydon, J.D. Sau, S. Tewari, and S. Das Sarma, Sci. Rep. {\bf 5}, 8880 (2015).
\bibitem{fu2009}
L. Fu and C. L. Kane, Phys. Rev. Lett. {\bf 102}, 216403 (2009).
\bibitem{akhmerov2009}
A. R. Akhmerov, J. Nilsson, and C. W. J. Beenakker, Phys. Rev. Lett. {\bf 102}, 216404 (2009).
\bibitem{law2009}
K. T. Law, P. A. Lee, and T. K. Ng, Phys. Rev. Lett. {\bf 103}, 237001 (2009).

\bibitem{mourik2012} 
V. Mourik, K. Zuo, S. M. Frolov, S. R. Plissard, E. P. A. M. Bakkers, and L.P. Kouwenhoven, Science {\bf 336}, 1003 (2012).
\bibitem{deng2012}
M. T. Deng, C. L. Yu, G. Y. Huang, M. Larsson, P. Caroff, and H. Q. Xu, Nano Lett. {\bf 12}, 6414 (2012).
\bibitem{finck2013}
A. D. K. Finck, D. J. Van Harlingen, P. K. Mohseni, K. Jung, and X. Li, Phys. Rev. Lett. {\bf 110}, 126406 (2013).

\bibitem{nadj-perge2014}
S. Nadj-Perge, I. K. Drozdov, J. Li, H. Chen, S. Jeon, J. Seo, A. H. MacDonald, B. A. Bernevig, and A. Yazdani, Science {\bf 346}, 602 (2014).

\bibitem{xu2015} 
J.-P. Xu, M.-X. Wang, Z.-L. Liu, J.-F. Ge, X. Yang, C. Liu, Z.-A. Xu, D. Guan, C.-L. Gao, D. Qian, Y. Liu, Q.-H. Wang, F.-C. Zhang, Q.-K. Xue, and J.-F. Jia, Phys. Rev. Lett. {\bf 114}, 017001 (2015).
\bibitem{chen2009}
Y. L. Chen, J. G. Analytis, J. H. Chu, Z. K. Liu, S. K. Mo, X. L. Qi, H. J. Zhang, D. H. Lu, X. Dai, Z. Fang, S. C. Zhang, I. R. Fisher, Z. Hussain, and Z. X. Shen, Science {\bf 325}, 178 (2009).

\bibitem{gygi1991}
F. Gygi and M. Sch\"ulter, Phys. Rev. B {\bf 43}, 7609 (1991).
\bibitem{mizushima2010}
T. Mizushima and K. Machida, Phys. Rev. A {\bf 82}, 023624 (2010).

\bibitem{bode2003}
M. Bode, Rep. Prog. Phys. {\bf 66}, 523 (2003).
\bibitem{hanaguri2014}
Y.-S. Fu, M. Kawamura, K. Igarashi, H. Takagi, T. Hanaguri, and T. Sasagawa, Nat. Phys. {\bf 10}, 815 (2014).
\end{thebibliography}
\end{document}